\documentclass[prb,
twocolumn,
superscriptaddress,showpacs,amsmath,amssymb,longbibliography]{revtex4-1}

\usepackage{xcolor}
\usepackage{graphicx}
\usepackage{hyperref}

\begin{document}

\title{From elasticity tetrads to rectangular vielbein}

\author{G.E.~Volovik}
\affiliation{Low Temperature Laboratory, Aalto University,  P.O. Box 15100, FI-00076 Aalto, Finland}
\affiliation{Landau Institute for Theoretical Physics, acad. Semyonov av., 1a, 142432,
Chernogolovka, Russia}

\date{\today}

\begin{abstract}
The paper is devoted to the memory of Igor E. Dzyaloshinsky. In our common paper I.E. Dzyaloshinskii and G.E. Volovick, 
Poisson brackets in  condensed matter, Ann. Phys.  {\bf 125} 67--97 (1980), we discussed the elasticity theory described in terms of the gravitational field variables -- the elasticity vielbein $E_\mu^a$. They come from the phase fields, which describe the deformations of crystal.  The important property of the elasticity vielbein $E^a_\mu$ is that in general they are  not the square mstrices. While the spacetime index $\mu$ takes the values $\mu=(0,1,2,3)$,  in crystals the index $a=(1,2,3)$,  in vortex lattices  $a=(1,2)$, and in smectic liquid crystals there is only one phase field, $a=1$. These phase fields can be considered as the spin gauge fields, which are similar to the gauge fields in Standard Model  (SM) or in Grand Unification (GUT). 

On the other hand, the rectangular vielbein $e^\mu_a$ may emerge in the vicinity of Dirac points in Dirac materials. In particular, in the planar phase of the spin-triplet superfluid $^3$He the spacetime index $\mu=(0,1,2,3)$, while the spin index $a$ takes values  $a=(0,1,2,3,4)$.
Although these $(4 \times 5)$ vielbein  describing the Dirac fermions are rectangular, the effective metric $g^{\mu\nu}$of Dirac quasiparticles  remains (3+1)-dimensional.
All this suggests the possible extension of the Einstein-Cartan gravity by introducing the rectangular vielbein,  where the spin fields belong to the higher groups, which may include SM or even GUT groups.

\end{abstract}
\pacs{
}

\maketitle 

\section{Introduction}

The paper Ref. \cite{Volovik2020b} was published in the JETP issue dedicated to Dzyaloshinskii's 90th birthday.
In that paper I discussed the application of the so-called elasticity vielbein, which were introduced in our common paper \cite{DzyalVol1980}, to gravity. 

The important property of the elasticity vielbein is that they have dimension of inverse length. If the gravitational tetrads have also this property, then the metric field $g_{\mu\nu}$, which is the bilinear combination of the tetrad fields, acquires the dimension of  the inverse square of length $[g_{\mu\nu}]=1/[l]^2$.  This is distinct from the conventional dimensionless metric, $[g_{\mu\nu}]=1$, in general relativity.  The same dimension of the metric field, $[g_{\mu\nu}]=1/[l]^2$, appears in the Diakonov theory \cite{Diakonov2011,VladimirovDiakonov2012,VladimirovDiakonov2014,ObukhovHehl2012},
in which the tetrad itself is the bilinear combination of the fermionic fields, and thus the metric is the quadruplet of the fermions.
The analog of the   Diakonov scenario takes place in one of topological phases of of superfluid $^3$He -- the B-phase \cite{Volovik1990}. 

Due to the change of the dimensionality of the metric,  the physical quantities, which obey diffeomorphism invariance,  become dimensionless. This takes place for such quantities as Newton constant, the scalar curvature, the cosmological constant, particle masses, fermionic and scalar bosonic fields, etc. 
This does not require the existence of the fundamental length, such as Planck length scale. The vacuum can be considered as superplastic (malleable) medium \cite{KlinkhamerVolovik2019}, in which the size of the elementary cell can be made arbitrary small.

Here I will discuss another interesting property of the elasticity vielbein. As distinct from the conventional gravitation tetrads, the 
elasticity vielbein is not necessary the square matrix. I will show several examples of the rectangular vielbein emerging in different systems, including Weyl and Dirac materials. When applied to gravity, this suggests that the gravity for fermions and bosons can be essentially different.
While the fermions can be described by the $4\times k$ vielbein matrix $e^\mu_a$ with $k \geq 4$, the bosonic fields see only the conventional $4\times 4$ metric  $g_{\mu\nu}$, which is the bilinear form of these rectangular vielbein fields \cite{Volovik2020}.

\section{Rectangular vielbein in elasticity theory}

Elasticity vielbein is a convenient way to discuss non-linear elastic deformations of crystals in terms of the curved space \cite{DzyalVol1980}
(the analog gravity in the elasticity theory of crystals see also in Refs.\cite{Bilby1955,Bilby1956,Kroener1960,AndreevKagan84,KleinertZaanen2004,HehlObukhov2007}). 
In the three-dimensional crystals, the vielbein $E^a_\mu$ with $a=1,2,3$ form three elastic $U(1)$ gauge fields, which reflect the periodicity in crystals
and three sets of crystal planes. 
As the gauge fields they have dimension of the inverse length,
which is different from the dimensionless conventional tetrads in general relativity.
The vielbein  $E^a_\mu$ provide the proper description of different anomalous topological phenomena in topological insulators \cite{NissinenVolovik2019,Vishwanath2019,NissinenHeikkila2021,Nissinen2020}. In particular, being the gauge fields, they together with the electromagnetic $U(1)$ gauge field $A_\mu$ enter the mixed topological Chern-Simons terms in the action.

The deformed crystal structure of the 3d crystal can be described as a system of three crystallographic surfaces of constant phase $X^a(x)=2\pi n^a$, $n^a \in {Z}$ with $a=1,2,3$. The intersection of the surfaces
\begin{equation}
X^1({\bf r},t)=2\pi n^1 \,\,, \,\,  X^2({\bf r},t)=2\pi n^2 \,\,, \,\, X^3({\bf r},t)=2\pi n^3 \,,
\label{points}
\end{equation}
are the nodes of a deformed crystal lattice. The elasticity vielbein are the gradients of the three phase functions:
\begin{equation}
E^{~a}_\mu(x)= \partial_\mu X^a(x)\,.
\label{reciprocal}
\end{equation}
In an equilibrium crystal lattice the quantities $E^a_i$ are lattice vectors of the reciprocal Bravais lattice.  In a deformed crystal, but in the absence of dislocations the vielbein $E^a_\mu$ satisfy the integrability condition:
\begin{equation}
\nabla_\nu E^a_\mu - \nabla_\mu E^a_\nu=0\,,
\label{intergrability}
\end{equation}
i.e. their field strength $F^a_{\mu\nu}=0$. The nonzero field strength -- the torsion field -- appears in the presence of dislocations.

Note that the  vielbein $E^a_\mu$ are non-square. While the space-time index $\mu$ is the $3+1$-dimensional, $\mu=0,1,2,3$, the phase space index  $a$ takes values $a=1,2,3$. This gives rise  to the rectangular $4\times 3$ vielbein  $E^a_\mu$. 

The rectangular vielbein appears also in the dynamics of the other crystalline structures, for example in dynamics of the vortex lattice in superfluids and superconductors. The vortex lattice has two phase functions $X^1({\bf r},t)=2\pi n^1$ and $X^2({\bf r},t)=2\pi n^2$, corresponding to two sets of crystal planes, and is described by  the $4\times 2$  vielbein \cite{VolovikDotsenko1979}. The dynamics of the smectic liquid crystals (see e.g. Ref. \cite{Kleman1975} and of the vortex sheets \cite{Volovik2015}, which have  single set of planes, is described by single phase function $X^1({\bf r},t)=2\pi n^1$ and thus by the $4\times 1$  vielbein. The periodically driven system (the Floquet crystal) serves as an exclusion: the vielbein becomes quadratic, forming the $4\times 4$ tetrads \cite{NissinenVolovik2018b}.

\section{Dirac fermions}

\subsection{Weyl fermions and  tetrads}
\label{WeylTetradsSec}

The quadratic $4\times 4$ vielbein -- tetrads -- emerge in the vicinity of the Weyl points  in Weyl semimetals and Weyl superfluids/superconductors \cite{Volovik1987,Horava2005}.
The reason for that is that the Weyl point is the topological object, which can be represented as a Dirac magnetic monopole in the momentum space -- a source or sink of Berry flux \cite{Volovik1987}.
In the vicinity of the monopole, the inverse Green's function of fermions can be written in terms of tetrads $e^\mu_a$:
\begin{equation}
G^{-1}(p_\mu) = \gamma^a e^\mu_a p_\mu \,,
\label{GreenWeyl}
\end{equation}
where the $\gamma$-matrices are:
\begin{eqnarray}
 \gamma^0 =1  \,\,,\,\, a=0  \,,
\label{gammaWeyl1}
\\
\gamma^a = \sigma^a   \,\,,\,\, a=1,2,3 \,.
\label{gammaWeyl2}
\end{eqnarray}
Here $p_0$ is the (quasi)particle energy; $p_i$ is the momentum counted from the Weyl point; and $\sigma^a$  for $a=1,2,3$ are the Pauli matrices, which correspond to the effective spin in the two-band structure. 

In the Weyl materials, tetrads $e^\mu_a$ serve as the analog of the gravitational tetrads. The  energy spectrum of the massless (gapless) fermions can be expressed in terms of the effective metric field \cite{NissinenVolovik2017}:
\begin{eqnarray}
g^{\mu\nu}p_\mu p_\nu=0\,,
\label{SpectrumWeyl1}
\end{eqnarray}
where the metric $g^{\mu\nu}$ is the bilinear combination of the tetrads:
\begin{eqnarray}
g^{\mu\nu}= e^\mu_ae^\nu_b \eta^{ab}\,, 
\label{SpectrumWeyl2}
\\
\eta^{ab}={\rm diag}(-1,1,1,1)\,.  
\label{SpectrumWeyl3}
\end{eqnarray}

Here the role of the phase space is played by the space of the effective spin. 
This phase space has the same dimension as the (orbital) coordinate spacetime. Both are governed by the Lorentz groups, 
$SO(1,3)_{\rm S}$ for the spin space and $SO(1,3)_{\rm O}$ for the spacetime. 

Note also that while the  vielbein
$E^a_\mu$, which emerges in the elasticity theory,  has the  covariant spacetime index $\mu$, the  tetrads $e^\mu_a$ emerging in the Dirac materials are the contravariant vectors. This demonstrates different origin of the tetrads in these two systems: geometric and dynamical. The elasticity tetrads $E^a_\mu$ are related to the geometry of crystal planes in coordinate space, while the Weyl tetrads are related to the topology in momentum space and thus to dynamics of Weyl fermions. 

\subsection{One-dimensional Dirac fermions with rectangular vielbein}

The non-square vielbein  $e^\mu_a$ can be obtained by decreasing the spacetime dimension without decreasing the spin space. For example, in the $1+1$ dimensional spacetime the inverse Green's function in Eq.(\ref{GreenWeyl}) is expressed in terms of the $2\times 4$ vielbein $e^\mu_a$:
\begin{equation}
G^{-1}(p_0,p_x)  = \gamma^a e^\mu_a p_\mu= \gamma^a (e^0_a p_0  +e^1_a p_x)\,,
\label{GreenFS}
\end{equation}
The  energy spectrum of these massless fermions can be expressed in terms of the metric field:
\begin{eqnarray}
g^{\mu\nu}p_\mu p_\nu=0 \,\,, \,\, \mu,\nu=(0,1)\,, 
\label{SpectrumFS1}
\\
g^{\mu\nu}= e^\mu_ae^\nu_b \eta^{ab}\,\,, \,\, a,b=(0,1,2,3)\,, 
\label{SpectrumFS2}
\\
\eta^{ab}={\rm diag}(-1,1,1,1)\,.  
\label{SpectrumFS3}
\end{eqnarray}
Eq.(\ref{GreenFS}) describes the 1+1 dimensional Dirac point (or in the other language the 0-dimensional Fermi surface). In the simple example, when 
$e^0_1=e^0_2=e^0_3=e^1_0=0$, one obtains the energy spectrum:
\begin{eqnarray}
 p_0 = \pm  v_Fp_x \,,
\label{1Dspectrum1}
\\
 v_F=\frac{|{\bf e}^1|}{e^0_0} \,\,,\,\,  {\bf e}^1=(e^1_1,e^1_2,e^1_3)\,.
 \label{1Dspectrum2}
\end{eqnarray}
Here  $v_F$ is the Fermi velocity, and the 3d spin vector ${\bf e}^1$ determines the axis of spin quantization.

The relativistic Dirac spectrum in Eq.(\ref{1Dspectrum1}) does not depend on the number of components of the vector ${\bf e}^1$, or in general it does not depend on the dimension of the phase (spin) space. Moreover, when the spectrum of Dirac fermions is expressed in terms of the effective metric in Eq.(\ref{SpectrumFS1}),  the dimension of spin space is lost. The $(1+1)$-dimensional metric $g^{\mu\nu}$, which describes the Dirac  fermions, has no information on the (3+1) dimension of the spin space. 

Another property of the rectangular vielbein concerns the metric $g_{\mu\nu}$, which enters the effective interval
\begin{equation}
 ds^2=g_{\mu\nu}dx^\mu dx^\nu \,.
\label{Interval}
\end{equation}
This metric is obtained as inverse to $g^{\mu\nu}$, but it cannot be obtained from the inverse vielbein, since the non-square vielbein  matrix $e^\mu_a$ is not invertible. Whether the Moore-Penrose
generalization of the inverse for rectangular matrices \cite{Penrose1955,Greville1966} (pseudoinverse) can be useful for the physics of vielbein  is an interesting question.

All this suggests that the same can be applied to Dirac and Weyl fermions in (3+1)-dimensional spacetime. Instead of  the $4\times 4$ tetrads  in the Sec. \ref{WeylTetradsSec}, the  Dirac and Weyl fermions can be described by the rectangular  $4\times k$ vielbein with $k>4$. Dirac fermions with $4\times 5$ vielbein emerge in the planar phase of the spin-triplet $p$-wave superfluids/superconductors \cite{Volovik2020}, which we shall discuss in the Section \ref{PlanarSec}.

\section{Rectangular vielbein  in planar phase of superfluid $^3$He}
\label{PlanarSec}

In the general spin triplet $p$-wave pairing state, the $2\times 2$ matrix of the gap function is:
\begin{equation}
 \hat{\Delta}({\bf p})=  A_{\alpha}^i\sigma^\alpha p_i \,,
\label{SpinTriplet}
\end{equation}
where $\sigma^\alpha$ are the Pauli matrices for spin and $A_\alpha^i$  is the $3\times 3$ complex matrix of the order parameter, see the book \cite{Vollhardt1990}.

The Bogoliubov-Nambu Hamiltonian for quasiparticles is the $4\times 4$ matrix, which includes the conventional spin  with 
$\boldmath{\mathrm{\sigma}}$-matrices and the Bogoliubov-Nambu particle-hole isospin with $\boldmath{\mathrm{\tau}}$-matrices:
\begin{equation}
  \begin{pmatrix}
  \epsilon(p) & \hat\Delta  \\
  \hat\Delta^\dagger & -\epsilon(p)
  \end{pmatrix}
   \,,
\label{HamiltonianGeneral}
\end{equation}
where $\epsilon(p)=c_\parallel(p-p_F)$, $c_\parallel=v_F$, and $v_F$ and $p_F$ are correspondingly the Fermi velocity and Fermi momentum of the normal Fermi liquid.

In  the planar phase the general form of the order parameter is:
\begin{eqnarray}
A_\alpha^i=c_\perp e^{i\Phi}\left(R_\alpha^i-   {\hat l}^i {\hat s}_\alpha \right) \,.
\label{OP1}
\end{eqnarray}
Here $\Phi$ is the phase of the order parameter, which we shall ignore later; $\hat{\bf l}$ is the unit vector of the orbital anisotropy of the planar phase; $R_\alpha^i$ is the matrix of rotation from the orbital space to spin space; ${\hat s}_\alpha = R_{\alpha k} {\hat l}^k$  is the unit vector of the uniaxial anisotropy in the spin space; and $p_Fc_\perp$ is the gap amplitude.

The quasiparticle spectrum has two Dirac points at ${\bf p}={\bf p}_{\rm Dirac}=\pm p_F\hat{\bf l}$.
To write the spectrum in the vicinity of the Dirac points, let us introduce the Dirac $\gamma$-matrices in the following way:
\begin{equation}
\gamma^0=i\tau_2 \,\,, \,\gamma^a=\tau_3\sigma^a \,\, (a=1,2,3)\,\,, \,    \gamma^4=\tau_1 \,,
\label{5gamma}
\end{equation}
where $\sigma^a$ and $\tau_a$ with  $a=1,2,3$ are two sets of Pauli matrices.
Five $\gamma$-matrices in Eq.(\ref{5gamma}) obey anticommutation relations:
\begin{equation}
\{\gamma^a,\gamma^b \}=2\eta^{ab}\,\,, \,    \eta^{ab}={\rm diag}(-1,1,1,1,1) \,.
\label{5gammaAnti}
\end{equation}

The properly determined inverse Green's function near one of the two Dirac points has the form of Eq.(\ref{GreenWeyl}): 
\begin{equation}
G^{-1}(p_\mu) = \gamma^a e^\mu_a p_\mu \,,
\label{GreenWeyl2}
\end{equation}
where the momentum $p_\mu$ is counted from the chosen Dirac point. 
The Green's function contains $4\times 5$ vielbein. The nonzero vielbein components in the planar state in Eq.(\ref{OP1}) at $\Phi=0$ are
 \begin{eqnarray}
 e^0_0=1 \label{e00} \,,
 \\
 e^i_a= c_\perp \left(R_\alpha^i-   {\hat l}^i {\hat s}_\alpha \right)\,\,\,  {\rm for} \,\, a=1,2,3\,,
\label{e0i}
\\
 e^i_4=c_\parallel \hat l^i\,.
\label{e4i}
\end{eqnarray}

In spite of the $(4\times 5)$-dimensional vielbein describing these fermions, their energy spectrum is given by the conventional 3+1 metric in Eq.(\ref{SpectrumWeyl1}):
\begin{eqnarray}
g^{\mu\nu}p_\mu p_\nu=0\,.
\label{SpectrumDirac1}
\end{eqnarray}
This metric is expressed in terms of  $(4\times 5)$ vielbein:
\begin{eqnarray}
g^{\mu\nu}= e^\mu_ae^\nu_b \eta^{ab} \,\,,\, \eta^{ab}={\rm diag}(-1,1,1,1,1)\,.
\label{SpectrumDirac2}
\end{eqnarray}
 
With vielbein in Eqs.(\ref{e00})-{(\ref{e4i}) the metric elements are:
  \begin{equation}
g^{00}=-1 \,\,,\,\, g^{ik}= c_\parallel^2 \hat l^i  \hat l^k + c_\perp^2 (\delta^{ik}-\hat l^i  \hat l^k)\,,
\label{3Dmetric2}
\end{equation}
  \begin{equation}
g_{00}=-1  \,\,,\,\,  g_{ik}= \frac{1}{c_\parallel^2} \hat l_i  \hat l_k + \frac{1}{c_\perp^2} (\delta_{ik}-\hat l_i  \hat l_k)\,.
\label{InverseMetric}
\end{equation}

It is interesting, that for $c_\parallel^2 =c_\perp^2\equiv c^2$ the metric corresponds to the Minkowski vacuum, $g_{00}=-1$ and $g_{ik}=(1/c^2)\delta_{ik}$,  even when the vielbein in Eqs.(\ref{e00})-{(\ref{e4i}) depends on space and time. In particular, the vielbein may have the topologically nontrivial structure. For example, it may have the global monopole configuration of the $\hat{\bf l}$-vector, $\hat{\bf l}({\bf r})=\hat{\bf r}$ \cite{Volovik2020}, but the space-time remains Minkowski (except for the region of the core of the monopole, where $c(r=0)=0$).
This is distinct from the conventional global monopoles in general relativity, which give rise to the conical spacetime
\cite{Starobinskii1977,Barriola1989,Monopole1990,Monopole2001,Bronnikov2002,Mavromatos2017,Petrov2019,Monopole2020,Kaloper2022}.

\section{Massive Dirac field with rectangular vielbein}

The massless Dirac fermions with $(4\times 5)$-dimensional vielbein can be extended to the massive case.
Dirac fermions with $(4\times 5)$ vielbein may arise in the Dirac materials when there are two successive phase transitions. The first transition produces the spin triplet $p$-wave phase corresponding to the planar phase in Sec. \ref{PlanarSec}. In the second transition the $Z_2$ time reversal symmetry is broken, and the imaginary $s$-wave component appears in the gap function. Such $p+is$ state was also discussed in superconductors, see e.g. Refs. \cite{Smidman2017,Shang2020,BitanRoy2020}. In the simplest case the planar phase order parameter in Eq.(\ref{OP1}) is extended to
\begin{eqnarray}
 \hat{\Delta}= e^{i\Phi} \left(c_\perp \left(R_\alpha^i - {\hat s}_\alpha  {\hat l}^i \right)p_i \sigma^\alpha + i \chi \right)\,.
\label{Planar+is2}
\end{eqnarray}
Here $\chi$ is the nonzero imaginary part of the $s$-wave component -- the real scalar Higgs field describing the broken time reversal symmetry.
On the Higgs fields in different superfluid and superconducting systems see review \cite{VolovikZubkov2014}.

This  $planar+is$ phase has massive Dirac fermions. Near the points ${\bf p}_{\rm Dirac}=\pm p_F\hat{\bf l}$ the proper inverse Green's function is
\begin{equation}
G^{-1}(p_\mu) = i\gamma^a e^\mu_a p_\mu +M\,.
\label{GreenDiracMass}
\end{equation}
Here the Dirac mass is $M=\chi$, while the $(4\times 5)$-dimensional vielbein $e^\mu_a$ remains the same as in the massless case, with $\mu=(0,1,2,3)$ and $a=(0,1,2,3,4)$.
The spectrum of massive Dirac particles is also determined by the same relativistic 3+1 metric:
\begin{equation}
g^{\mu\nu}p_\mu p_\nu+M^2=0\,.
\label{SpectrumDiracMass}
\end{equation}
The $planar+is$ phase has the topologically stable global monopole in the same manner as in the pure planar phase. It is also the hedgehog in the $\hat{\bf l}$-vector with $\hat{\bf l}({\bf r})=\pm \hat{\bf r}$, and this monopole also  does not disturb the metric if $c_\parallel^2 =c_\perp^2\equiv c^2$.  Such objects are impossible for the conventional Dirac, but they become possible, if the Dirac matrices or vielbein have nontrivial topology, i.e. the nontrivial dependence on spacetime. This suggests that the nontrivial topology of the $\gamma$-matrices can be the proper extension of general relativity.

Note the difference with the conventional Dirac equation. In the latter case the matrix $\gamma^4$ is actually the $\gamma^5$ matrix, which is responsible for the axial mass in the mass  term $M_1 + M_2\gamma^5$. In our case, $\gamma^4$ enters the kinetic  term  $\gamma^4e_4^\mu p_\mu$, and the spin  group is $SO(1,4)_{\rm S}$.  The group  $SO(1,4)_{\rm S}$ can be also obtained from the Lorentz group by extension of the discrete symmetry to the continuous symmetry \cite{MaiezzaNesti2022} (how the discrete group $Z_2$ transforms to the continuous group $U(1)$ in the planar phase, is demonstrated in  
Ref. \cite{Makhlin2014}).

\section{Possible extension to GUT}

The example of the planar phase suggests that $4\times 4$ Dirac $\gamma$-matrices can be extended to the matrices with higher spin dimension without extension of the space dimension. In general, one may have the Lorentz
$SO(1,N)_{\rm S}$ group for spin space and the Lorentz $SO(1,d)_{\rm O}$ group for (orbital) spacetime with $N> d$,
and with $\Gamma$ matrices obeying the Clifford algebra
\begin{equation}
\{\Gamma^a,\Gamma^b \}=2\eta^{ab}\,.
\label{GammaAnti}
\end{equation}
This results in the rectangular $(d+1)\times (N+1)$ vielbein $e^\mu_a$, which enter the Green's function:
\begin{equation}
G^{-1}(p_\mu) = i\Gamma^a e^\mu_a p_\mu +M\,,
\label{GreenDiracGeneral}
\end{equation}
while the metric is the square $(d+1)\times (d+1)$ matrix:
\begin{eqnarray}
g^{\mu\nu}= e^\mu_ae^\nu_b \eta^{ab}\,.
\label{MetricGeneral}
\end{eqnarray}
This metric enters the conventional energy spectrum of massive particle in the $d+1$ spacetime:
\begin{equation}
g^{\mu\nu}p_\mu p_\nu+M^2=0\,.
\label{SpectrumGeneral}
\end{equation}

 It is important that for $N>d$ the inverse vielbein are absent.
The spin and spacetime indices are raised and lowered with $\eta^{ab}$ and with metric $g^{\mu\nu}$ correspondingly.
But now the spin matrix $\eta_{ab}$ cannot have expression similar to Eq.(\ref{MetricGeneral}) for metric:
 \begin{equation}
\eta_{ab} \neq g_{\mu\nu}e^\mu_a e^\nu_b \,.
\label{tildeeta}
\end{equation}
This can be checked multiplying both sides by $\eta^{ab}$:
\begin{equation}
\eta^{ab}\eta_{ab}=N+1 \neq  \eta^{ab}g_{\mu\nu}e^\mu_a e^\nu_b =g_{\mu\nu}g^{\mu\nu} = d+1 \,.
\label{Trace}
\end{equation}

Another property of the rectangular vielbein with $N>d$ concerns the metric determinant.
For $N=d$, the  metric determinant is always negative, since $g=-e^2<0$, where $e$ is the tetrad determinant. For the non-square vielbein matrix the determinant is absent, and thus there is no constraint on the sign of the metric determinant. 
Let us consider the simplest case of the $4\times 5$ vielbein, when the $4\times 4$ tetrad part of the vielbein is diagonal, while the the extra vielbein part $e^\mu_4$ contains only nonzero element $e^0_4$.
Then the metric elements are
\begin{equation}
g^{ik}=\delta^{ik} \,\,,\,\, g^{00}=- (e^0_0)^2 +(e^0_4)^2 
\,,
\label{g00}
\end{equation}
which allows for the signature change.

As a result, the phase transition with the change of the sign of the determinant becomes possible. In this type of the tetrad gravity the metric  signature becomes dynamical. The discussion of the dynamical metric signature can be found in Refs. \cite{Bondarenko2021,BondarenkoZubkov2022}.

The extension of the spin space from $1+3$ to $N+3$ with $N>3$ is essentially different from the extension of the spacetime. As distinct from string theories, the coordinate spacetime remains $3+1$ dimensional. The spin space extension has something common with the unification of GUT with Lorentz group \cite{MaiezzaNesti2022}. Example is the group $SO(1,11)$, which corresponds to $N=11$. This group can be decomposed into the subgroups:
 \begin{equation}
SO(1,11) \sim  SO(10)  \times SO_F(6) \times SO(1,3)_{\rm S} 
\,.
\label{11}
\end{equation} 
Here $SO(10)$ is the GUT group; $SO_F(6)=SU_F(4)$ is the family group for 4 generations of fermions; and  $SO(1,3)_{\rm S}$ is the Lorentz spin  group. The number of generators of $SO(1,11)$ group is $66=45+15+6$.
The Clifford algebra of the  corresponding  ${\Gamma}$-matrices is Cl$_{11,1}({\bf R})_{\bf C}$.
See also the IKKT (Ishibashi-Kawai-Kitazawa-Tsuchiya \cite{IKKT1997}) matrix model in string theory (or the type IIB matrix model) \cite{Nishimura2022}, where there is the $SO(1,9)$ group with 10-dimensional  $\Gamma$-matrices, and 16 fermionic variables.

In principle, the reduction from the  $(d+1) \times (N+1)$ vielbein to $(d+1) \times (d+1)$ vielbein may correspond to the symmetry breaking (topological) phase transition   $SO(1,N)\rightarrow SO(1,d)$.

\section{Conclusion}

The elasticity vielbein discussed in Refs. \cite{DzyalVol1980,NissinenVolovik2019} come from the phase fields, which describe the deformations of crystal. These phase fields can be considered as the spin gauge fields, which are similar to the gauge fields in Standard Model or in Grand Unification. That is why the  spin fields can be incorporated into the higher group, which may include both the spin group and the groups of SM or GUT.
The same concerns the extension of the Lorentz spin group for Dirac fermions to the higher group, which may also include the SM or GUT groups.

Here we considered several examples of such extension, which demonstrated that the gravitational tetrads can be generalized to the non-square (rectangular) vielben, while the metric remains (3+1)-dimensional. This happens for the spin-triplet $p$-wave planar phase, where the Dirac quasiparticles are described by the $(4 \times 5)$-vielbein, but by the $(4 \times 4)$ effective metric.
The possible extension of the GUT theory is considered, which incorporates the $SO(1,3)_{\rm S}$ spin group.

How the rectangular vielben modifies the Einstein-Cartan equations of general relativity, can be the interesting problem.
This especially concerns the spin connection and torsion fields.

{\bf Acknowledgements}. 
I thank Jaakko Nissinen for discussions. This work has been supported by the European Research Council (ERC) under the European Union's Horizon 2020 research and innovation programme (Grant Agreement No. 694248).

.


\begin{thebibliography}{15}


\bibitem{Volovik2020b}
G.E. Volovik,
Dimensionless physics,
ZhETF {\bf 159}, 815--821 (2021),
JETP {\bf 132},  727--733 (2021),
arXiv:2006.16821.

\bibitem{DzyalVol1980}
I.E. Dzyaloshinskii, and G.E. Volovick, 
Poisson brackets in  condensed matter,
Ann. Phys.  {\bf 125} 67--97 (1980).

 \bibitem{Diakonov2011}
 D. Diakonov,
 Towards lattice-regularized Quantum Gravity,
 arXiv:1109.0091.

 \bibitem{VladimirovDiakonov2012}
A.A. Vladimirov and D. Diakonov,
Phase transitions in spinor quantum gravity on a lattice,
Phys. Rev. D {\bf 86}, 104019 (2012).

 \bibitem{VladimirovDiakonov2014}
A.A. Vladimirov and D. Diakonov,
Diffeomorphism-invariant lattice actions,
Physics of Particles and Nuclei {\bf 45}, 800 (2014).

 \bibitem{ObukhovHehl2012}
Y.N. Obukhov and F.W. Hehl,
Extended Einstein–Cartan theory a la Diakonov: The field equations,
Phys. Lett. B {\bf 713}, 321--325 (2012).

\bibitem{Volovik1990}
G.E. Volovik, 
Superfluid $^3$He-B and gravity, 
Physica B {\bf 162}, 222--230 (1990).

 \bibitem{KlinkhamerVolovik2019}
F.R. Klinkhamer and G.E. Volovik,
Tetrads and $q$-theory,
Pis'ma ZhETF  {\bf 109}, 369--370 (2019),
 JETP Lett. {\bf 109},  362--365 (2019),
arXiv:1812.07046.

\bibitem{Volovik2020}
G.E. Volovik,
Vielbein with mixed dimensions and gravitational global monopole in the planar phase of superfluid $^3$He,
Pis’ma v ZhETF {\bf 112},  539--540  (2020),
JETP Lett. {\bf 112},  505--507 (2020),
arXiv:2009.09779.

\bibitem{Bilby1955} 
 B.A. Bilby and E. Smith,
 Continuous distributions of dislocations: A new application of the methods of non-Riemannian geometry, 
 Proc. Roy. Soc. Sect. A {\bf 231}, 263--273 (1955).
 
 \bibitem{Bilby1956} 
 B.A. Bilby and E. Smith, 
Continuous distributions of dislocations. III, 
 Proc. Roy. Soc. Sect. A {\bf 236}, 481--505 (1956).
 
 \bibitem{Kroener1960}
 E. Kr\"oner,
Allgemeine Kontinuumstheorie der Versetzongen and Ligenspannunge,
 Arch. Rational Mech. Anal. {\bf 4}, 18--334 (1960).

\bibitem{AndreevKagan84}
A.F. Andreev, M. Yu. Kagan,
Hydrodynamics of a rotating superfluid liquid,
Zh. Eksp. Teor. Fiz. {\bf 86},546 (1984) [Sov. Phys. JETP {\bf 59} (1984)]

\bibitem{KleinertZaanen2004} 
H. Kleinert and J. Zaanen, 
World nematic crystal model of gravity explaining the absence of torsion, 
Phys. Lett. A {\bf 324}, 361--365 (2004).

\bibitem{HehlObukhov2007} 
F.W. Hehl and Y.N. Obukhov,
Elie Cartan’s torsion in geometry and in field theory, an essay,
Annales de la Fondation Louis de Broglie {\bf 32}, 157 --194 (2007).

\bibitem{NissinenVolovik2019}
J. Nissinen and G.E. Volovik,
Elasticity tetrads, mixed axial-gravitational anomalies, and (3+1)-d quantum Hall effect,
Physical Review Research {\bf 1}, 023007 (2019),
arXiv:1812.03175.

\bibitem{Vishwanath2019}
Xue-Yang Song, Yin-Chen He, Ashvin Vishwanath, and Chong Wang,
Electric polarization as a nonquantized topological response and boundary Luttinger
theorem,
arXiv:1909.08637.

\bibitem{NissinenHeikkila2021}
J. Nissinen, T.T. Heikkil\"a and G.E. Volovik,
Topological polarization, dual invariants, and surface flat band in crystalline insulators,
Phys. Rev. B {\bf 103}, 245115 (2021),
arXiv:2008.02158.

\bibitem{Nissinen2020}
J. Nissinen,
Field theory of higher-order topological crystalline response, generalized global symmetries and elasticity tetrads,
arXiv:2009.14184 .

\bibitem{VolovikDotsenko1979}
G.E. Volovik and Vik.S. Dotsenko, 
Poisson brackets and continual dynamics of the vortex lattice in rotating HeII,
Pisma ZhETF {\bf 29}, 630--633 (1978),
JETP Lett. {\bf 29} 576--579 (1978)..

\bibitem{Kleman1975}
M. Kleman and O. Parodi, 
Covariant elasticity for smectics A, 
J. Phys. {\bf 36}, 671--681 (1975).

\bibitem{Volovik2015}
G.E. Volovik,
Superfluids in rotation: Landau-Lifshitz vortex sheets vs Onsager-Feynman vortices,
Physics Uspekhi {\bf 58}, 897--905 (2015),
arXiv:1504.00336.

\bibitem{NissinenVolovik2018b}
J. Nissinen and G.E. Volovik,
Tetrads in solids: from elasticity theory to topological quantum Hall systems and Weyl fermions,
ZhETF {\bf 154},   1051--1056 (2018),
arXiv:1803.09234.

\bibitem{Volovik1987}
G.E. Volovik,  
Zeros in the fermion spectrum in superfluid systems as diabolical points,
Pisma ZhETF {\bf 46}, 81--84 (1987),
JETP Lett. {\bf 46}, 98--102 (1987).

\bibitem{Horava2005}
P. Horava,
Stability of Fermi surfaces and K-theory,
Phys. Rev. Lett. {\bf 95}, 016405 (2005).

\bibitem{NissinenVolovik2017}
J. Nissinen and G.E. Volovik,
Type-III and IV interacting Weyl points, 
JETP Lett.  {\bf 105},  447--452 (2017),
arXiv:1702.04624.

\bibitem{Penrose1955}
R. Penrose, 
A generalized inverse for matrices, 
Proc. Cambridge Philos. Soc., {\bf 51},406--413 (1955).

\bibitem{Greville1966}
T.N.E. Greville, 
Note on the generalized inverse of a matrix product,
SIAM Review {\bf 8}, 518--521 (1966).


\bibitem{Vollhardt1990}
D. Vollhardt and P. W\"olfle, 
{\it The Superfluid Phases of Helium 3},
 (Taylor \& Francis, London, 1990).

\bibitem{Starobinskii1977}
D.D. Sokoloff, A.A. Starobinskii,	
On the structure of curvature tensor on conical singularities,
Dokl. Akad. Nauk SSSR {\bf 234}, 1043--1046 (1977),
Sov. Phys. Dokl. {\bf 22}, 312 (1977).

\bibitem{Barriola1989}
M. Barriola and A. Vilenkin,
Gravitational field of a global monopole,
Phys. Rev. Lett. {\bf 63}, 341-343 (1989).

\bibitem{Monopole1990}
D. Harari and C. Loustó,
Repulsive gravitational effects of global monopoles,
Phys. Rev. D {\bf 42}, 2626 (1990).

\bibitem{Monopole2001}
E.R. Bezerra de Mello,
Physics in the global monopole spacetime,
 Brazilian Journal of Physics  {\bf 31}, 211 (2001).
   
   
\bibitem{Bronnikov2002}
K. A. Bronnikov, B. E. Meierovichc and E. R. Podolyak,
Global monopole in general relativity,
JETP {\bf 95},  392--403 (2002).

\bibitem{Mavromatos2017}
N.E. Mavromatos and S. Sarkar,
Magnetic monopoles from global monopoles in the presence of a Kalb-Ramond field,
Phys. Rev. D {\bf 95}, 104025 (2017).

\bibitem{Petrov2019}
J. R. Nascimento, Gonzalo J. Olmo, P. J. Porfírio, A. Yu. Petrov and A. R. Soares,
Global monopole in Palatini $f({\cal R})$ gravity,
Phys. Rev.  D {\bf 99}, 064053 (2019).

\bibitem{Monopole2020}
E.A.F. Braganca, R.L.L. Vitoria, H. Belich, E.R. Bezerra de Mello,
Relativistic quantum oscillators in the global monopole spacetime,
Eur. Phys. J. C {\bf 80}, 206 (2020).

\bibitem{Kaloper2022}
N. Kaloper,
Troubles with global monopoles in quantum gravity,
International Journal of Modern Physics D {\bf 31}, 2250035 (2022).


\bibitem{Smidman2017}
M. Smidman, M.B. Salamon, H.Q. Yuan and D.F. Agterberg,
Superconductivity and spin-orbit coupling in non-centrosymmetric materials: a review,
Rep. Prog. Phys. {\bf 80},  036501 (2017).

\bibitem{Shang2020}
T. Shang,
Simultaneous nodal superconductivity and time-reversal symmetry breaking in the noncentrosymmetric superconductor CaPtAs,
Phys. Rev. Lett. {\bf 124}, 207001 (2020).

\bibitem{BitanRoy2020}
B. Roy,
Higher-order topological superconductors in P-, T -odd quadrupolar Dirac materials,
Phys. Rev. B {\bf 101}, 220506(R) (2020),
arXiv:2003.12566.

\bibitem{VolovikZubkov2014}
G.E. Volovik and M.A. Zubkov,
Higgs bosons in particle physics and in condensed matter,
J. Low Temp. Phys. {\bf 175}, 486--497 (2014):
arXiv:1305.7219.


\bibitem{MaiezzaNesti2022}
A. Maiezza and F. Nesti,
Parity from gauge symmetry,
arXiv:2111.11076.

\bibitem{Makhlin2014}
Yu. Makhlin, M. Silaev and G.E. Volovik,
Topology of the planar phase of superfluid $^3$He and bulk-boundary correspondence for three-dimensional topological superconductors,
Phys. Rev. B {\bf 89} 174502 (2014);
arXiv:1312.2677.

\bibitem{Bondarenko2021}
S. Bondarenko,
Dynamical signature: complex manifolds, gauge fields and non-flat tangent space,
arXiv:2111.06095

\bibitem{BondarenkoZubkov2022}
S. Bondarenko and M.A. Zubkov,
Riemann-Cartan gravity with dynamical signature,
Pis’ma v ZhETF {\bf 116}, \#1 (2022),
JETP Lett. {\bf 116}, \#1 (2022).


\bibitem{Volovik2003b}
 G.E. Volovik, 
 Dark matter from $SU(4)$ model, 
Pisma ZhETF {\bf 78}, 1203--1206 (2003),
JETP Lett. {\bf 78},  691--694 (2003),
hep-ph/0310006.


\bibitem{IKKT1997}
N. Ishibashi, H. Kawai, Y. Kitazawa and A. Tsuchiya,
A large-$N$ reduced model as superstring,
Nuclear Physics B {\bf 498},  467--491 (1997).

\bibitem{Nishimura2022}
J. Nishimura,
Signature change of the emergent space-time in the IKKT matrix model,
arXiv:2205.04726.


\end{thebibliography}
\end{document}